\documentclass[5p,times,letterpaper]{elsarticle}
\usepackage[english]{babel}
\usepackage{graphicx} % Required for inserting images
\usepackage{float}
\usepackage{tikz}
\usepackage{subcaption}

\usepackage{amsmath}
\usepackage{booktabs}

\usepackage[breaklinks]{hyperref}
\hypersetup{hidelinks}
\usepackage{url}

\usepackage{parskip}
\usepackage{blindtext}
\usepackage{balance}
\usepackage{multicol}

\usepackage{csquotes}

\usepackage{emptypage}

\journal{Nuclear Engineering and Technology}

\begin{document}
\sloppy
\begin{frontmatter}
\title{Extending OpenMC Validation to Spent Fuel Canisters: A Criticality Benchmark Against MCNP}
\author[add1]{J. Ruiz-Pineda\corref{cor2}} 
\author[add1,add2]{J. Romero-Barrientos\corref{cor1}}
\ead{jaime.romero@cchen.cl}

\author[add1,add2,add3]{F. Molina\corref{cor3}}
\author[add1]{M.~Zambra\corref{cor4}}
\author[add1,add2,add3]{F.~L\'opez-Usquiano\corref{cor5}}
\cortext[cor1]{Corresponding author}
\address[add1]{Centro de Investigación en F\'isica Nuclear y Espectroscop\'ia de Neutrones CEFNEN, Comisi\'on Chilena de Energ\'ia Nuclear, Nueva Bilbao 12501, Las Condes, Santiago, Chile}
\address[add2]{Millennium Institute for Subatomic physics at high energy frontier - SAPHIR, Fern\'andez Concha 700, Las Condes, Santiago, Chile}
\address[add3]{Departamento de Ciencias Físicas, Universidad Andres Bello, Sazi\'e 2212, 837-0136, Santiago, Chile}

% --------------------------------------------------------------------------------------------
\begin{abstract}
OpenMC is an open-source Monte Carlo code with increasing relevance in criticality safety and reactor physics applications. While its validation has covered a broad range of systems, its performance in spent nuclear fuel storage scenarios remains limited in the literature. This work benchmarks OpenMC against MCNP for eleven configurations based on the KBS-$3$ disposal concept, involving variations in geometry, fuel composition (fresh vs spent), and environmental conditions (e.g., air, argon, flooding scenarios). Effective multiplication factors (k-eff) and leakage fractions were evaluated for both codes. Results show strong agreement, with code-to-code k-eff differences below $0.8 \%$ in dry storage conditions, and consistent trends across all cases. Notably, OpenMC successfully captures inter-canister neutron interaction effects under periodic boundary conditions, demonstrating its applicability to dry storage configurations. This benchmark supports the extension of the validation domain of OpenMC toward SNF transport and disposal applications.
\end{abstract}
\end{frontmatter}
\section{Introduction} %-----------------------------------------------------------------------
The world annual reactor-related uranium requirements are projected to increase from $60,114$~tU/year in $2020$ to between $63,040$ and $108,272$~tU/year by $2040$~\cite{NEA_uranium2022}. This increase in uranium demand will lead to a corresponding rise in spent nuclear fuel (SNF) inventories over the coming decades. Addressing the challenges of SNF management, the International Atomic Energy Agency (IAEA) has highlighted the critical need for developing public acceptance, long-term funding, and disposal facilities for high-level waste (HLW) and SNF when considered as waste~\cite{IAEA_status}. Given this scenario, research and development on techniques related to SNF management will become increasingly relevant.

After being used in the reactor, nuclear fuel is removed from the reactor core and can be handled following two main types of fuel cycles: (i) the closed fuel cycle, in which spent fuel is considered to provide a potential future energy resource; and (ii) the open fuel cycle, in which spent fuel is considered as waste~\cite{IAEA_status}. In the open fuel cycle option, spent fuel is stored for several decades to allow the decay heat to be reduced. After a period of storage, the spent fuel is encapsulated in a robust and corrosion-resistant container to meet disposal acceptance criteria and is then disposed in a deep geological repository (DGR)~\cite{IAEA_status}.

A disposal method, defined as the intentional emplacement of radioactive waste in a facility without the intent to retrieve~\cite{IAEA_status} (i.e., permanent storage), is widely accepted among technical experts as the preferred method for ensuring the long-term safety of SNF and HLW. The development of deep geological repositories (DGRs) has been the focus of the disposal strategies of many countries~\cite{IAEA_status}.

\begin{figure}
    \centering
    \includegraphics[width=\linewidth]{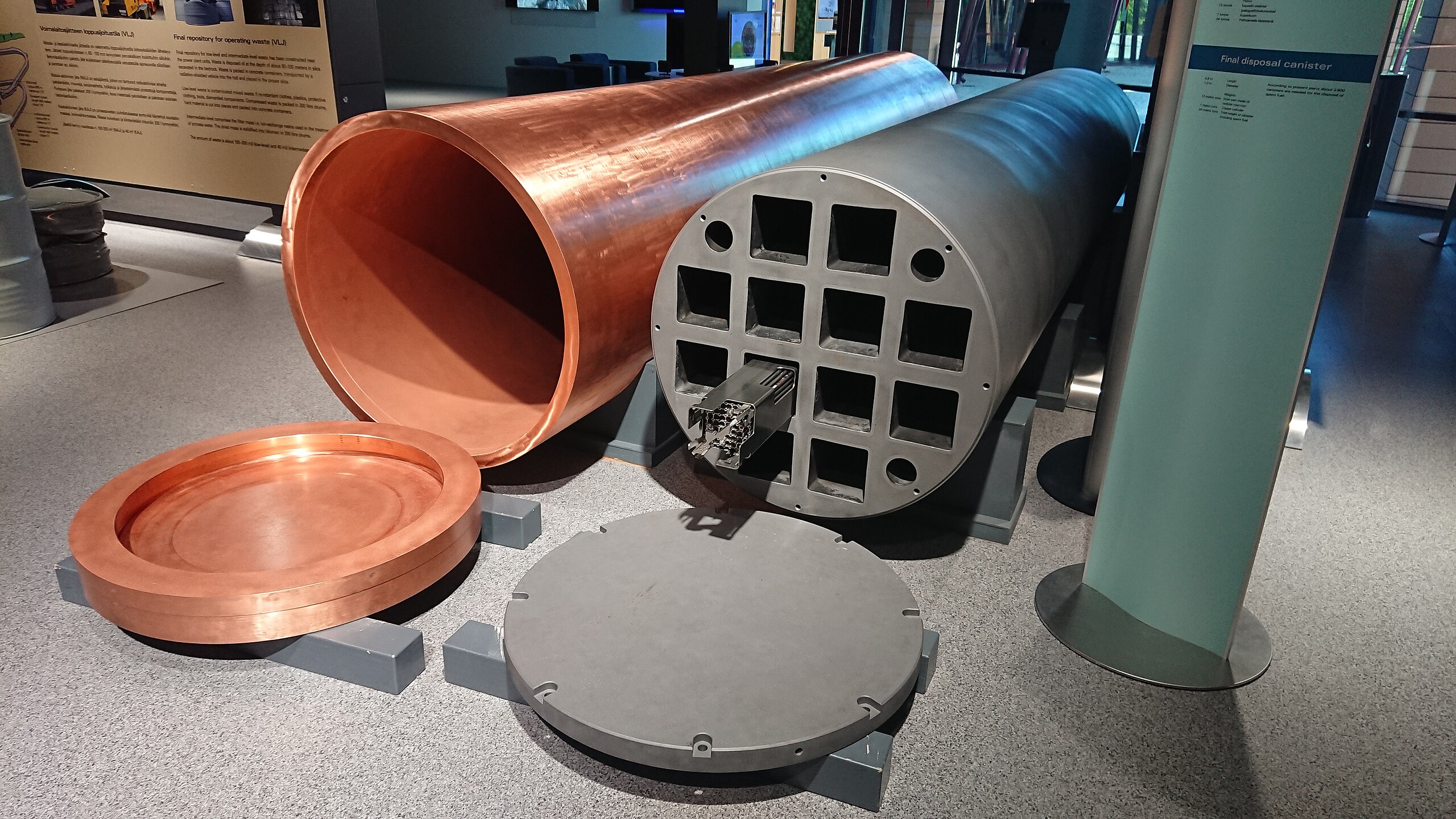}
    \caption{Olkiluoto final disposal canister (\href{https://commons.wikimedia.org/wiki/File:Olkiluoto-final-disposal-canister.jpg}{Teemu Väisänen}, \href{https://creativecommons.org/licenses/by-sa/4.0/legalcode}{CC BY-SA $4.0$}). On the left, a copper overpack; on the right, a cast iron insert. In this work, a canister with PWR dimensions was considered for the simulations.}
    \label{fig:foto_canister}
\end{figure}

In the $1970$s, Swedish power utilities formed the SKB Nuclear Fuel Safety Project, which developed a disposal method compatible with a DGR called KBS (\textit{Kärnbränsle Sakerhet}, Nuclear Fuel Safety in English), where canisters are placed in deposition tunnels at a depth of about $420$ meters below the ground surface into individual vertical deposition holes, referred to as KBS-$3$V (V for vertical)~\cite{posiva_thermal}. The original design included a canister with a lead basket for storing fuel rods extracted from spent fuel assemblies, encased in a copper overpack~\cite{KBS2_1978}. Later, the lead insert was replaced by a cast iron insert, and fuel assemblies are now disposed as a unit~\cite{KBS3_2010}. This approach has influenced the designs of the Finnish company Posiva Oy~\cite{posiva_canister}, which has made significant progress in implementing the KBS-$3$ method (the modern design is shown in Figure~\ref{fig:foto_canister}).

A criticality safety evaluation (CSE) is defined as \enquote{the analysis and documentation that fissionable material processes remain subcritical under both normal and credible abnormal conditions}~\cite{DOE_standard}. Therefore, the effective multiplication factor ($k_{\mathit{eff}}$ or k-eff) will be a key parameter in this work, as our main comparison quantity. According to paragraph $6$.$6$ of the International Atomic Energy Agency (IAEA) safety guide SSG-$15$, \enquote{Only verified and validated methods should be used for predicting the safety of operational states or the consequences of accidents} in SNF storage facilities~\cite{IAEA_storage}. Thus, in this context, to conduct a CSE and use its results for safety predictions, it is mandatory to use a validated code (i.e. a computational predictive method).
Validation and verification of computational methods are typically based on the ICSBEP Handbook, which $2021$ version contains a set of $587$ criticality-related evaluations~\cite{ICSBEP_handbook}.
Studies related to SNF storage or disposal methods calculations have been conducted using codes such as MCNP~\cite{spentfuel_MCNP}, SCALE~\cite{spentfuel_SCALE}, Serpent~\cite{spentfuel_SERPENT}, TRIPOLI~\cite{spentfuel_TRIPOLI}, GEANT4~\cite{spentfuel_GEANT4} and MCkeff~\cite{spentfuel_MCkeff}. A study that compares criticality calculations of KENO\footnote{part of the \href{https://scale-manual.ornl.gov/Keno.html}{SCALE} Code System.} and OpenMC over decay time using a spent fuel cask has been done in the past with satisfactory results~\cite{KENO_OpenMC}, supporting the interest on the application of OpenMC on spent fuel disposal systems.
Within the variety of Monte Carlo codes, MCNP, SCALE, Serpent and TRIPOLI are subject to export control~\cite{licencias_montecarlo} regulations. Therefore, there is the need for an accessible Monte Carlo code for this kind of applications and OpenMC appears as a promising option.

\section{OpenMC Monte Carlo code} %--------------------------------------------------
OpenMC is an open source Monte Carlo particle transport code developed at MIT and published in $2015$. It is currently maintained and continuously developed by an international community. Originally published with a focus on neutron criticality calculations~\cite{openmc_romano}, OpenMC is currently capable of performing fixed source, k-eigenvalue, and subcritical multiplication calculations on models built using either a constructive solid geometry or CAD representation. In addition to these capabilities, OpenMC is able to programming pre- and post-processing, multi-group cross section generation, workflow automation, depletion calculations, multiphysics coupling, and the visualization of geometry and tally results. Nuclear data is handled in HDF5 format; and it can be run in parallel using a hybrid MPI and OpenMP programming model, which has been extensively tested on leadership class supercomputers \cite{openmc_homepage}.
Several validation and verification studies have been made, mainly in nuclear reactor systems~\cite{openmc_validations}, including a comparison with MCNP and experimental data for time dependent phenomena, obtaining satisfactory results (e.g., a maximum of 0.3\% difference between MCNP and OpenMC for the k-eff)~\cite{openmc_romero}. Nevertheless, at the moment of writing this article, no public criticality safety evaluations of spent nuclear fuel canisters using OpenMC were found in the literature.

\section{Methodology} %------------------------------------------------------------------------
\label{sec:methodology}
In this work, criticality calculations were performed for a spent nuclear fuel disposal configuration using OpenMC and MCNP, with the aim of exploring  the applicability of OpenMC to this type of system. By benchmarking against MCNP, this study offers supportive evidence for the use of OpenMC in spent fuel canister analyses — a domain where its application has been limited so far.
For this study, SNF canisters were simulated in a normal operation situation and in a set of cases that include an external water flood and/or leakage into the canister (described at section \ref{sec:env_specifications}). Both spent and fresh fuel were considered for all calculations, except for a case where the focus was on the boundary condition of the simulation rather than on the SNF behavior. For each case, the effective multiplication factor was calculated using OpenMC and MCNP. Throughout this study, objects, materials, and environments represent simplified yet representative geometries intended for benchmarking purposes.
To validate OpenMC results, the same simulations were performed using MCNP writing equivalent inputs (using the same materials, surfaces and cells definitions), the same nuclear cross section library, execution settings and boundary conditions. These parameters are shown in Table \ref{t:computation}. Since in MCNP~\cite{mcnp6_manual} implicit capture is activated by default, in OpenMC the same setting was activated. Interested readers can access the inputs of one of the cases through the.
\begin{table*}[tbhp]
    \centering
    \begin{tabular}{|c|c|c|}
        \hline
         & \textbf{OpenMC} & \textbf{MCNP} \\
        \hline
        Version & OpenMC(TD) Ver. 2.1~\cite{openmc_romero} & MCNP6, 1.0~\cite{mcnp6_release}\\
        \hline
        Cross sections library & ENDF/B-VIII.0~\cite{endfbviii0} & ENDF/B-VIII.0~\cite{endfbviii0}\\
        \hline
        Particles per batch / hist. per cycle & 50,000 & 50,000 \\
        \hline
        Batches / cycles & 3,000 & 3,000\\
        \hline
        Inactive batches / cycles & 300 & 300 \\
        \hline
        Initial source position\footnotemark & x,y,z = 0,0,1 cm & x,y,z = 0,0,1 cm\\
        \hline
        Initial source angular distribution & Isotropic & Isotropic\\
        \hline
        Boundary surfaces condition & Vacuum & Vacuum\\
        \hline
    \end{tabular}
    \caption{Input paramaters for the simulations executed in this work.}
    \label{t:computation}
\end{table*}
\footnotetext{Given that the coordinate system origin $(0,0,0)$ coincides with a defined surface in the geometry, the particle source was positioned at $(0,0,1)$~cm, applying a $1$~cm offset in the z-direction to avoid conflicts with the initial cell assignment during simulation initialization.}
To compare the results obtained with OpenMC with our reference -MCNP-, this work considered the regular difference \enquote*{Diff. (pcm)}, as shown in Equation \ref{eq:diff_pcm}; and percentual relative difference \enquote*{Diff. (\%)}, as shown in Equation \ref{eq:diff_per}.
\begin{equation}
    \text{Diff. (pcm)}=\left(k_\textit{OpenMC}-k_\textit{MCNP}\right)\cdot 10^{5}
    \label{eq:diff_pcm}
\end{equation}
\begin{equation}
    \text{Diff. (\%)}=\left(\frac{k_\textit{OpenMC}}{k_\textit{MCNP}}-1\right)\cdot 100
    \label{eq:diff_per}
\end{equation}
Where $k_\textit{OpenMC}$ and $k_\textit{MCNP}$ are the Effective Multiplication Factors obtained with OpenMC and MCNP, respectively.
\section{Systems description} %---------------------------------------------------------------
\subsection{Canister and fuel assemblies}
For the specifications of the SNF canister, geometries and materials were retrieved from a report by Posiva Oy~\cite{posiva_canister}, which includes three types of insert (an example is shown in Fig. \ref{fig:foto_canister}); for this work, the PWR type was used since corresponds to the most common kind of reactor intended for energy generation~\cite{genIV_handbook_pioro}.

\begin{figure}[ht]
    \centering
    \includegraphics[width=\linewidth]{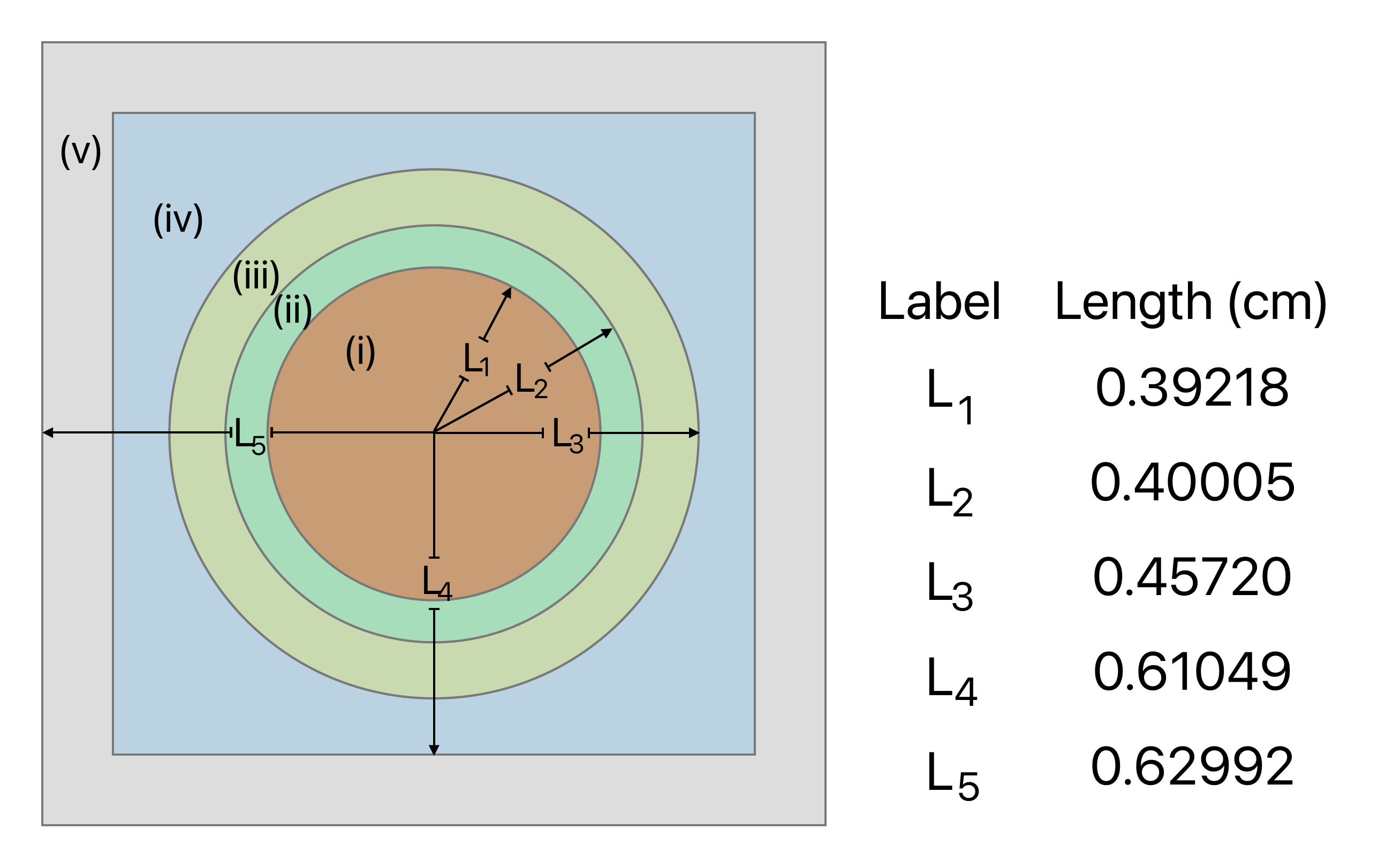}
    \caption{Radial cross-section of the fuel rod model used in the simulations. The regions labeled (i) to (v) correspond to: (i) fuel, (ii) gas gap, (iii) cladding, (iv) inner spacer boundary, and (v) outer spacer boundary. Distances L\_${}_{1}$ through L\_${}_{5}$ indicate radial positions from the center to the corresponding region interfaces. The table on the right lists the exact values of each radial dimension, expressed in centimeters. These geometric parameters define the fuel rod structure implemented in the Monte Carlo model. Note that the diagram is not to scale and is intended solely for schematic illustration.~\cite{BEAVRS}.}
    \label{fig:fuel_rod_xy}
\end{figure}

Regarding the geometric specifications of the fuel assemblies, a simplification of the PWR assemblies described on the BEAVRS benchmark~\cite{BEAVRS} was considered. The radial and axial geometry of the fuel rods and the radial geometry of the spacers were kept. In this study, $17\times17$ spaces assemblies were considered (with all $289$ spaces filled with fuel rods), each rod containing a $380.416$ cm long fuel cylinder with the radial geometry shown in Figure \ref{fig:fuel_rod_xy}. Radially, the fuel rod as an unit consist of a series of layers, the most internal is a solid piece of fuel, followed by a thin layer of fission-produced gas, an M5 cladding that seals the fuel, an intermediate space filled with argon and an M5 frame that sets the border to the neighboring rod.

The nozzles are one of the main parts of the support and alignment structures, they also stay with the fuel rods at the final disposal. In order to emulate the parts that composes the nozzles (i.e. springs, support legs, screws, and others), an homogeneous mixture of $50$ v.\%\footnote{in this work v.\% is the volume proportion.} steel and argon was made, similar to the proposal from BEAVRS~\cite{BEAVRS}.
Internal spaces in the fuel bundles and gaps within the canister are filled with argon by default in this work, this is corresponding with designs from Posiva and it is due to advantageous thermal and chemical properties from this gas ~\cite{posiva_canister}\cite{posiva_thermal}, exceptions are described in the further subsection.

The OpenMC and MCNP input files used in this study, including geometry, materials, and simulation parameters for the  \textit{Normal Operation – Spent Fuel} configuration, have been made publicly available through Zenodo~\cite{ruiz-pineda2025input}. This aims to promote transparency and reproducibility of the benchmark results.

In this work, two canister environment sets were considered for the canister location: \textit{Normal operation} and \textit{Flood}; each of them is detailed in the next subsection.

\subsection{Canister environments and specifications\label{sec:env_specifications}}
In criticality safety studies, boundary conditions are often selected to yield conservative estimates of the multiplication factor, with periodic or reflective surfaces commonly used to overestimate k-eff. In this work, vacuum boundary conditions were applied to the \textit{Normal Operation} (NO) and \textit{Flood} (FLD) scenarios, except in one case where periodic boundaries were tested. While vacuum conditions may underestimate k-eff due to particle loss at boundaries, this effect is expected to be minimal in large systems like the NO environment, where most particles are absorbed before reaching the system boundary. This consideration is relevant for interpreting potential differences between codes, particularly in configurations where neutron leakage is non-negligible.

In OpenMC, the leakage fraction provides a measure of how many particles escape the geometry without undergoing further interactions. A near-zero leakage value indicates that particles are predominantly terminated by nuclear reactions (e.g., absorption), while higher leakage suggests potential underestimation of the multiplication factor due to boundary losses. For the NO environment, which features large boundaries, leakage is expected to be negligible. In contrast, the FLD cases — which represent more compact or hypothetical geometries — can exhibit non-negligible leakage. To assess its impact, an additional comparison between vacuum and periodic boundary conditions was performed.

\subsubsection{Normal operation environment}
Criticality calculations for these canisters made by Posiva locates the canister in a fitted box (considering that the canister is $1.05$ m wide), that either will have a $1.1\times1.1$ m$^2$ base for an isolated dry canister and periodic lattice array; or a $1.6\times1.6$ m$^2$ base for an isolated flooded canister~\cite{Anttila_2005_fit}. On the one hand, simple and fit environments similar to the ones considered in this work are common in criticality evaluations for SNF canisters~\cite{Calic_2012_fit}~\cite{Anttila_1999_fit}~\cite{Yun_Kim_Park_Hong_2016_fit}. On the other hand, wider environments have been used in other studies, considering simplified materials and structures, with sizes up to an estimated $6\times6\times10$ m$^3$ box~\cite{Ibrahim_2022_wide}~\cite{Yun_Han_Cho_2022_wide}~\cite{Kim_Kim_Lee_Na_Chung_Kim_2022_wide}.

The NO environment considers a deep geological repository scenario, which consists on a concrete coated tunnel filled with air and a vertical hole to locate the canister filled with bentonite and other construction materials~\cite{Ibrahim_2022_wide}. Outside these structures there is an upper crust rock filling. The geometry for this environment is based on thermal analyses that contain more detail on the structures surrounding the canister~\cite{posiva_thermal}~\cite{Ibrahim_2022_wide}, obtaining a high detail $10\times10\times20$ m$^3$ environment shown in Figures \ref{fig:opnormal_xz} and \ref{fig:opnormal_xy}.

\subsubsection{Flood}
\begin{figure*}[hbt]
    \centering
    % Left side for the rectangular image as a subfigure
    \begin{minipage}{0.29\textwidth}
        \centering
        \begin{subfigure}{\textwidth}
            \centering
            \includegraphics[width=\linewidth]{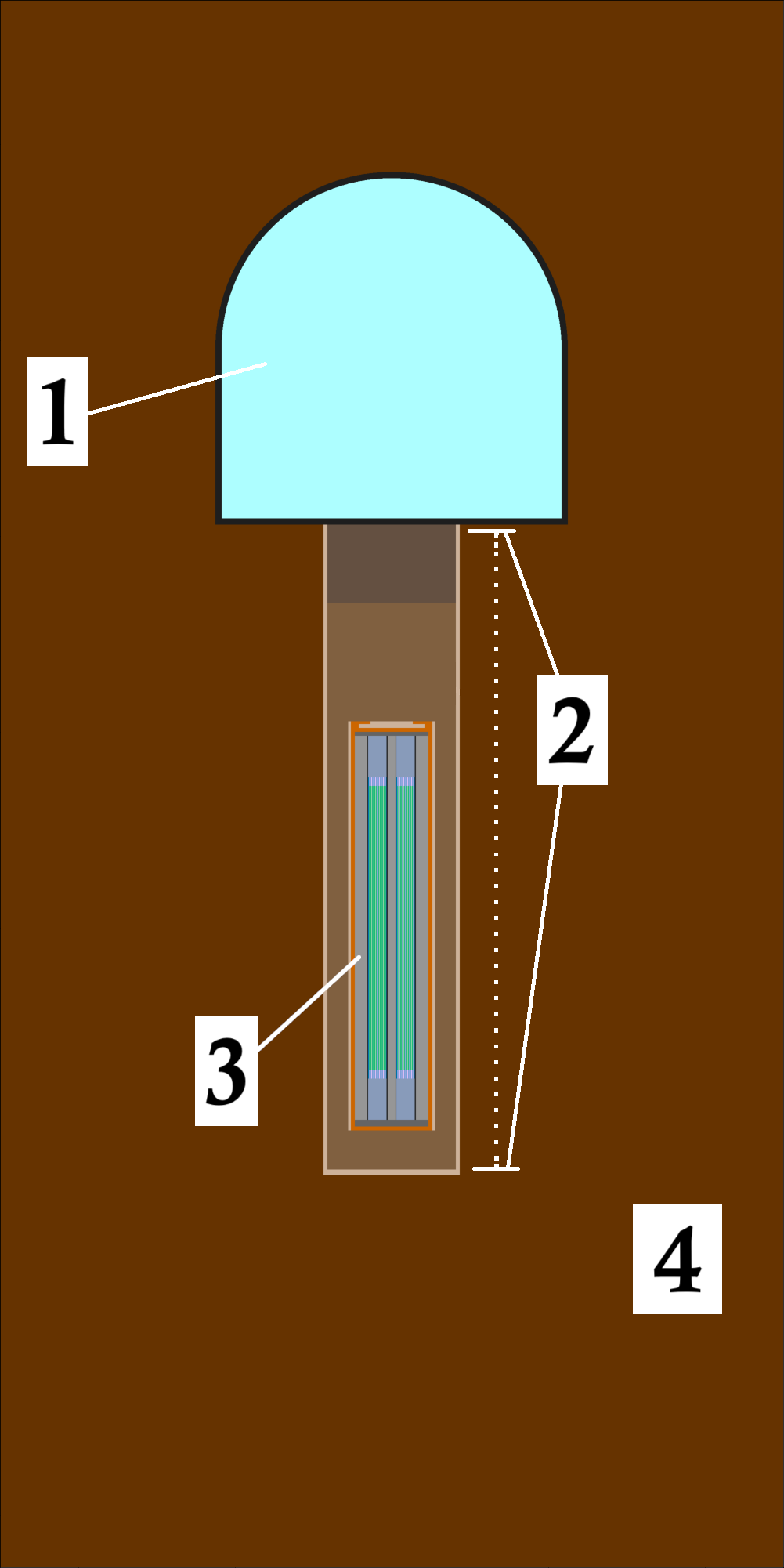}
            \caption{NO environment (Vertical view, full system). $1$: Tunnel. $2$: Borehole. $3$: Canister. $4$: Rock surrounding.}
            \label{fig:opnormal_xz}
        \end{subfigure}
    \end{minipage}%
    \hspace{1cm} % Custom horizontal space of 1cm between minipages
    % Right side for the two stacked square images as subfigures
    \begin{minipage}{0.23\textwidth}
        \centering
        % First square image
        \begin{subfigure}{\textwidth}
            \centering
            \includegraphics[width=\linewidth]{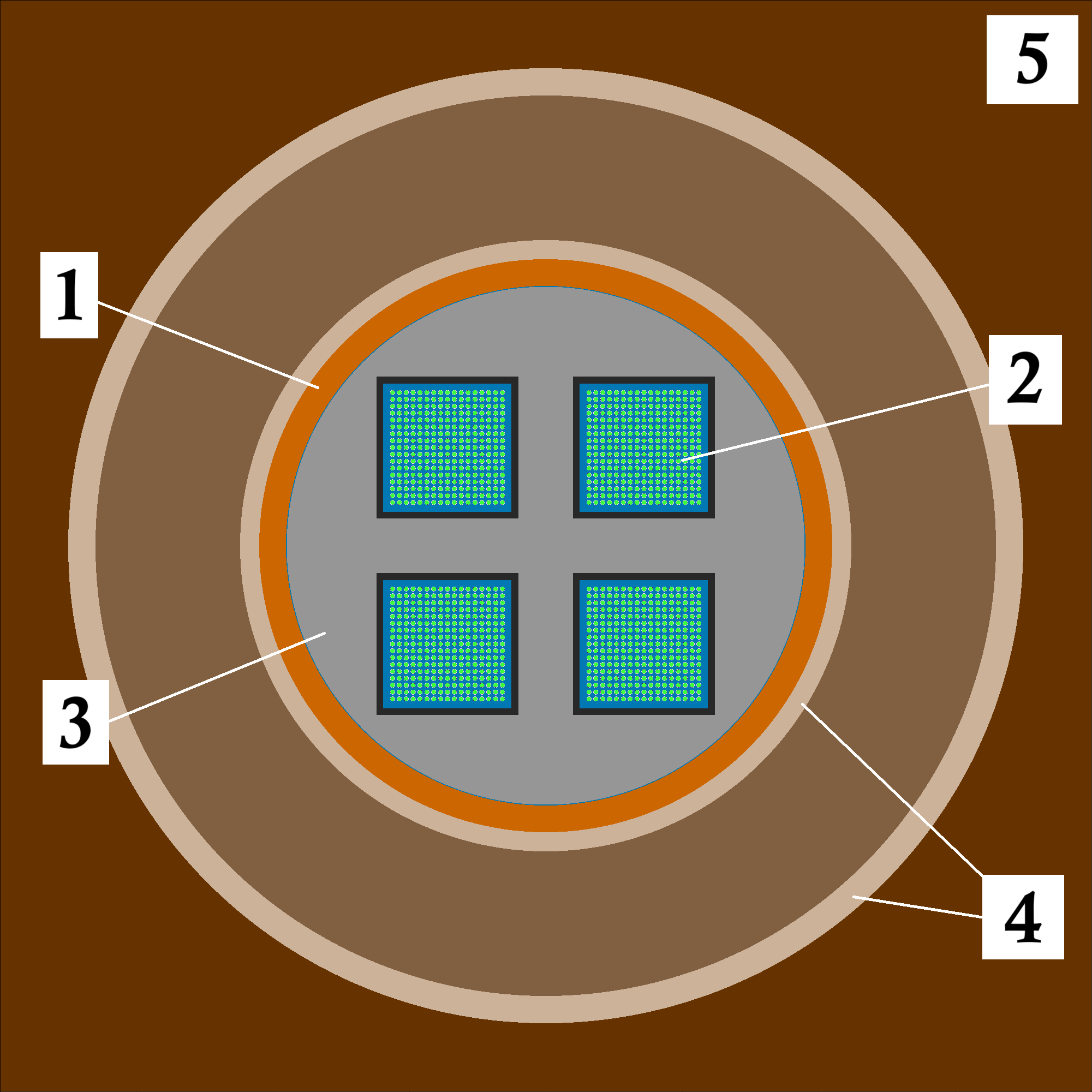}
            \caption{NO environment (Horizontal view, zoomed in). $1$: Copper overpack. $2$: Fuel assembly. $3$: Iron insert. $4$: Bentonite pellet layer. $5$: Rock surrounding.}
            \label{fig:opnormal_xy}
        \end{subfigure}
        \vfill
        % Second square image
        \begin{subfigure}{\textwidth}
            \centering
            \includegraphics[width=\linewidth]{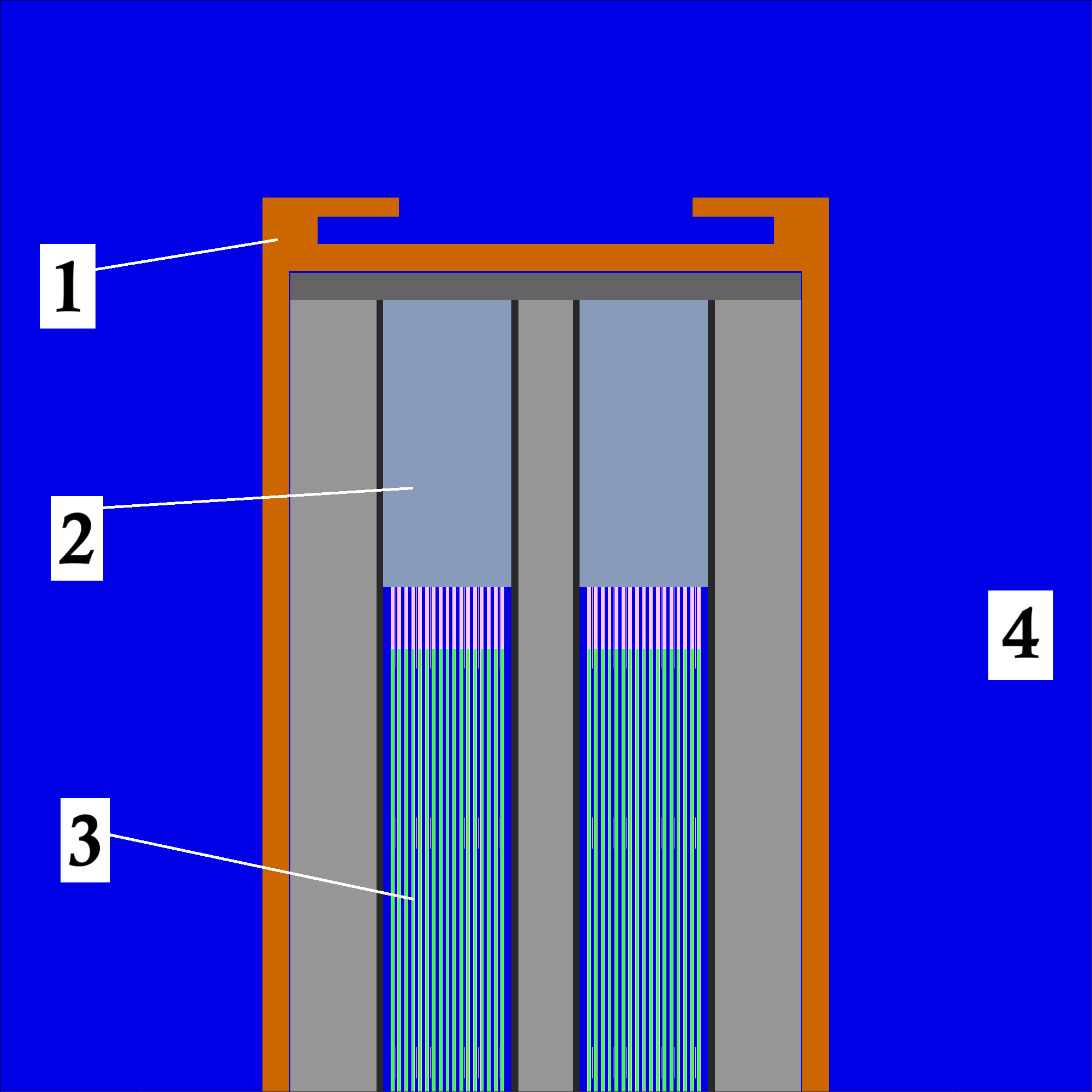}
            \caption{FLD environment, \enquote{Water in - Water out} case (Vertical view, zoomed in). $1$: Copper overpack. $2$: Assembly nozzles. $3$: Water surrounded fuel rods. $4$: Water surrounding.}
            \label{fig:flood}
        \end{subfigure}
    \end{minipage}
    \caption{Visualization of the main parts of the simulated systems. Made with the OpenMC plot tool.}
    \label{fig:overall_stacked}
\end{figure*}

In the FLD environments, the canister is located in a $5\times5\times10$ m$^3$ box that could be either filled with air or water. The insides of the canister can be either unchanged (argon) or emulating a water leakage, in the latter, the argon gas in the empty spaces inside the canister has been replaced by water and the nozzle material is now a mixture of steel and water. The presence of water through either an external flood or a leakage are considered as contingencies since the increase of hydrogenated materials near the SNF provokes an increase in neutron moderation through elastic scattering, and, therefore, an increase in the k-eff. These variations give rise to four distinct configurations, which in this study are classified based on the material inside the canister (\textit{in}) and the material filling the surrounding environment (\textit{out}). Figure \ref{fig:flood} illustrates the Water in – Water out geometry.

\subsubsection{Boundary condition change}
In addition to the previously described environmental configurations, this study also explored the impact of boundary conditions. When a particle reaches the external limits of the simulation geometry, a \textit{vacuum} boundary condition terminates its track and simulation. A \textit{periodic} boundary condition, on the other hand, transports the particle from the border surface to a \textit{pair} surface set by the user (usually at the opposite side of the system), keeping the particle within the geometry and allowing the continuation of its simulation.This algorithm effectively emulates the repetition of the system an indefinite number of times.
Several canisters placed in proximity may exhibit neutronic coupling due to inter-canister neutron transport. To examine how such coupling might influence simulation results, the \textit{Argon in – Air out} case with spent fuel was modeled using both vacuum and periodic boundary conditions. The difference in k-eff values between these two boundary types provides a reference for assessing the sensitivity of the system to inter-canister interactions. This scenario was chosen as a simplified dry storage configuration, and serves to explore how boundary conditions may affect k-eff in both OpenMC and MCNP.

\subsection{Materials}
Initially, the SNF isotopic specification was obtained from a benchmark PWR SNF composition intended as a starting point for decay calculations~\cite{spent_fuel_composition}, this composition contains $112$ nuclides with atomic weight fractions ranging from $2.73\times 10^{-21}$ to $6.72\times10^{-1}$.
For this work $1.5\times 10^{8}$ particles were simulated for each case studied, implying that, in the case that every particle interacts with the SNF, a minimum concentration of $\sim 0.67 \times 10^{-8}$ is necessary for a nuclide to have an average of interactions greater than $1$ per simulation. Therefore, it can be considered that nuclides with a concentration of $10^{-9}$ or less will not be object of interaction.
Since the main objective of this work is evaluating the OpenMC application on SNF criticality safety evaluations through a benchmark with MCNP, a set of justified simplifications was applied to the SNF composition, consistent with the goal of benchmarking OpenMC against MCNP in representative scenarios. As a simplification measure, three steps were applied: (i) nuclide concentrations were reduced by decay over a period of $30.5$ days ($\sim 1$~month), based on their half-lives~\cite{endfbviii0}; this duration was selected to represent a plausible timeframe for SNF transport and logistics, while omitting extended decay pool storage for modeling simplicity; (ii) only nuclides with a concentration fraction above $10^{-9}$ were retained, resulting in a set of $60$ nuclides—among which three (($^{14}$C, $^{129m}$Te and $^{242m}$Am) are not included in the ENDF/B-VIII.0 cross section library; and (iii) these three nuclides were excluded from the simulations, yielding a final spent fuel composition of $57$ nuclides.
For the composition of other materials, certain standards were considered -based on technical reports made by Posiva Oy~\cite{posiva_thermal}~\cite{posiva_canister}- which can be found in the third column of Table~\ref{t:materials}.%\balance 

\begin{table*}[phtb]
    \centering
    \small
    {\def\arraystretch{1.1}
    \begin{tabular}{@{}ccc@{}}\toprule
        \textbf{Part(s)} & \textbf{Material} & \textbf{Specifications and/or reference} \\
        \midrule
        Spent Fuel & Spent UO2 based fuel & initial $4.5$ wt.\% $^{235}$U, $50$ GWd/MTU~\cite{spent_fuel_composition} \\
        \hline
        Fresh fuel & Uranium dioxide & $4.5$ wt.\% $^{235}$U\\
        \hline
        Spent Inner rod gas & Helium &~\cite{BEAVRS}\\
        \hline
        Fresh Inner rod gas & Fission gas &~\cite{fuel_rod_examinations}\\
        \hline
        Fuel rod cladding \& spacers & M5 & with maximum impurity concentration~\cite{M5_specs}\\
        \hline
        Assembly empty space filling & Argon & natural abundance composition, $1$ atm~\cite{posiva_canister}\cite{hand_chem_phys}\\
        \hline
        Nozzle & Steel \& argon & S355JR EN10025-2:2004, $50$ v.\% each~\cite{prof_steel}\\
        \hline
        Profile & Steel & EN 10210 S355J2H ~\cite{prof_steel}\\
        \hline
        Insert & Cast iron &  EN(-1563)-GJS-400-15(GGG40)~\cite{cast_iron}\\
        \hline
        Insert lids & Steel & S355JR EN10025-2:2004~\cite{lid_steel}\\
        \hline
        Overpack &  Oxygen free copper & C102 standard~\cite{copper}\\
        \hline
        Filling pellet & Bentonite \& air & MX80 bentonite, $50$ v.\% air~\cite{bentonite}~\cite{hand_chem_phys}\\
        \hline
        Hole buffer & Bentonite & MX80 bentonite~\cite{bentonite}\\
        \hline
        Tunnel coating & Projected concrete & steel-reinforced~\cite{prof_steel}~\cite{sand_composition}~\cite{portland}~\cite{shotcrete}\\
        \hline
        Dense backfill & Bentonite \& cement & $50$ v.\% each~\cite{bentonite}~\cite{portland}\\
        \hline
        Tunnel filling & Air & $1.327$ wt.\% water~\cite{astro_quantities}\\
        \hline
        Outside of the tunnel & Continental crust rock &~\cite{continental_crust}\\
        \hline
        Flooding environment & Pure light water &~\cite{hand_chem_phys}\\
        \bottomrule
    \end{tabular}
    }
    \normalsize
    \caption{Materials considered in the simulations.}
    \label{t:materials}
\end{table*}
\section{Results and Discussion\label{sec:res}} %-------------------------------------------
In this section, the obtained effective multiplication factors will be presented and compared following Sec.~\ref{sec:methodology}. A comparison between results for both codes was performed on progressively more complex systems, from a fuel rod section to the full canister, using fresh fuel and argon canister filling on every system.

\subsection{Normal operation \label{ssec:res_normal}}
At this environment, differences between OpenMC and MCNP stayed below $0.8\%$ as shown in Table \ref{t:res_normal}. The replacement of spent fuel for fresh fuel translated into a $\sim46\%$ increase in k-eff for both codes. %Regardless this increase, the NO environment stays within the $\text{k-eff}=0.95$ and $\text{k-eff}=0.97$ safety limit for spent fuel and fresh fuel, respectively, since the presence of fresh fuel is classified as a contingency or abnormal condition in the context of radioactive waste canisters, meaning that the systems in this environment are not only subcritical but also comply with international safety requirements.
\begin{table*}[htpb]
    \centering
    \begin{tabular}{cccccc}
        \toprule
        \textbf{Case} & \textbf{k-eff OpenMC} & \textbf{k-eff MCNP} & \textbf{Diff. (pcm)} & \textbf{Diff. (\%)} & \textbf{Leakage}\\
        \midrule
        Spent Fuel & 0.19448(2) & 0.19587(2) & -139 & -0.71 &0.00001(2*)\\
        Fresh Fuel & 0.28310(2) & 0.28536(2) & -226 & -0.79 &0.00001(2*)\\
        \midrule
        Fuel diff. (\%) & +45.6 & +45.7 & & & \\
        \bottomrule
    \end{tabular}
    \caption{Comparison of k-eff values in the NO Environment for Spent and Fresh Fuel. Diff. (pcm) corresponds to the difference in pcm between OpenMC and MCNP Results. Leakage is the fraction of neutrons escaping the system boundaries. k-eff and leakage uncertainties are statistical uncertainties, while Diff. (pcm) and Diff (\%) are propagated uncertainties. [*]: [*] Leakage uncertainty reported as zero by OpenMC was replaced with a minimum value of 2 pcm, corresponding to the statistical resolution limit (1/50,000), to avoid misleading precision.  See Sec.~\ref{sec:sum_conc} for details.}
    \label{t:res_normal}
\end{table*}

\subsection{Flood}
On one hand, facing a water leakage, canisters with spent fuel responded with multiplication factors near $0.78$ and a code-to-code results difference of $0.56$\%. In systems without water leakage, k-eff values showed a $0.7$\% difference between results from both codes.
On the other hand, canisters with fresh fuel showed supercritical values (k-eff$>1$) in situations containing water leakage into the canister. For all cases with water filled boxes, the neutron leakage fraction resulted null within statistical boundaries.
%The effect of replacing spent fuel for fresh fuel are shown in Table \ref{t:res_flood_comparison}.
\begin{table*}[htpb]
    \centering
    \begin{tabular}{cccccc}
        \toprule
        \textbf{Case} & \textbf{k-eff OpenMC} & \textbf{k-eff MCNP} & \textbf{Diff. (pcm)} & \textbf{Diff. (\%)} & \textbf{Leakage}\\
        \midrule
         Water in - water out & $0.77731(6)$ & $0.78165(6)$ & $-434$ & $-0.56$ & $0.00000(2*)$\\
         Water in - air out & $0.77640(6)$ & $0.78075(6)$ & $-435$ & $-0.56$ & $0.07871(2)$\\
         Argon in - water out & $0.18334(2)$ & $0.18462(2)$ & $-128$ & $-0.69$ & $0.00000(2*)$\\
         Argon in - air out & $0.18092(2)$ & $0.18218(2)$ & $-126$ & $-0.69$ & $0.49964(4)$\\
         \bottomrule
    \end{tabular}
    \caption{Comparison of k-eff values in the FLD cases for Spent Fuel. Diff. (pcm) corresponds to the difference in pcm between OpenMC and MCNP Results. Leakage is the fraction of neutrons escaping the system boundaries. k-eff and leakage uncertainties are statistical uncertainties, while Diff. (pcm) and Diff (\%) are propagated uncertainties. The differences stay below $0.7 \%$. [*]: Leakage uncertainty reported as zero by OpenMC was replaced with a minimum value of 2 pcm, corresponding to the statistical resolution limit (1/50,000), to avoid misleading precision. See Sec.~\ref{sec:sum_conc} for details.}
    \label{t:res_flood_spent}
\end{table*}
\begin{table*}[htpb]
    \centering~
    \begin{tabular}{cccccc}
        \toprule
        \textbf{Case} & \textbf{k-eff OpenMC} & \textbf{k-eff MCNP} & \textbf{Diff. (pcm)} & \textbf{Diff. (\%)} & \textbf{Leakage}\\
        \midrule
         Water in - water out & $1.04877(7)$ & $1.05456(6)$ & $-579$ & $-0.549$ & $0.00000(2*)$\\
         Water in - air out & $1.04763(7)$ & $1.05336(6)$ & $-573$ & $-0.544$ & $0.07617(2)$\\
         Argon in - water out & $0.26340(2)$ & $0.26554(2)$ & $-214$ & $-0.81$ & $0.00000(2*)$\\
         Argon in - air out & $0.25923(2)$ & $0.26136(2)$ & $-213$ & $-0.81$ & $0.48803(3)$\\
         \bottomrule
    \end{tabular}
    \caption{Comparison of k-eff values in the FLD cases for Fresh Fuel. Diff. (pcm) corresponds to the difference in pcm between OpenMC and MCNP Results. Leakage is the fraction of neutrons escaping the system boundaries. k-eff and leakage uncertainties are statistical uncertainties, while Diff. (pcm) and Diff (\%) are propagated uncertainties. The differences stand up to $0.81~\%$. [*]: Leakage uncertainty reported as zero by OpenMC was replaced with a minimum value of 2 pcm, corresponding to the statistical resolution limit (1/50,000), to avoid misleading precision. See Sec.~\ref{sec:sum_conc} for details.}
    \label{t:res_flood_fresh}
\end{table*}
\begin{table*}[htpb]
    \centering
    \begin{tabular}{ccc}
        \toprule
        \textbf{Case} & \textbf{k-eff OpenMC (\%)} & \textbf{k-eff MCNP (\%)} \\
        \midrule~
         Water in - water out & $+34.9$ & $+34.9$ \\
         Water in - air out & $+34.9$ & $+34.9$ \\
         Argon in - water out & $+43.7$ & $+43.8$ \\
         Argon in - air out & $+43.3$ & $+43.5$ \\
         \bottomrule
    \end{tabular}
    \caption{Comparison of k-eff values in the FLD cases after replacement of spent fuel for fresh fuel. Shown as a percentual difference considering the spent fuel result as reference.}
    \label{t:res_flood_comparison}
\end{table*}

\subsection{Boundary type change}
To assess the potential for neutronic coupling between neighboring SNF canisters, a comparison was performed between vacuum and periodic boundary conditions using the configuration with the highest observed leakage fraction (i.e., Argon in – Air out). The k-eff increased by $+1.27(2)~\%$ in OpenMC and $+1.32(2)~\%$ in MCNP when using periodic boundaries, suggesting a consistent sensitivity in both codes to boundary-induced interaction effects. This confirms that canisters placed $3.95$~m apart in a simplified dry storage-like geometry may exhibit inter-canister neutron coupling. While this scenario does not represent a full physical model of dry storage, the results offer insight into how boundary conditions and geometry assumptions can influence criticality estimates across simulation tools.

\begin{table*}[htpb]
    \centering
    \begin{tabular}{ccccc}
        \toprule
        \textbf{Boundary type} & \textbf{k-eff OpenMC} & \textbf{k-eff MCNP} & \textbf{Diff. (pcm)} & \textbf{Diff. (\%)}\\
        \midrule
        Vacuum & $0.18092(2)$ & $0.18218(2)$ & $-126$ & $-0.69$ \\
        Periodic & $0.18321(2)$ & $0.18458(2)$ & $-137$ & $-0.74$ \\
        k-eff diff. (\%) & $+1.27(2)$ & $+1.32(2)$ &  &  \\
        \bottomrule
    \end{tabular}
    \caption{Comparison of k-eff values in the \enquote*{Argon in - Air out} Environment with Spent Fuel changing the boundary type. Diff. (pcm) corresponds to the difference in pcm between OpenMC and MCNP Results. keff uncertainties are statistical uncertainties, while Diff. (pcm) and Diff (\%) are propagated uncertainties. The differences stand to a maximum of $137$~pcm. The leakage for the periodic system resulted in $0.47077(4)$.}
    \label{t:res_boundary}
\end{table*}
\section{Summary \& Conclusions}\label{sec:sum_conc}%---------------------------------------------------------------------------
In this work, criticality calculations of systems containing KBS-$3$ spent nuclear fuel canisters were conducted, considering normal operation conditions, a set of abnormal conditions that associate two contingencies, an external flood and a water leakage into the canister, and a periodic lattice boundary condition to briefly explore potential canister-to-canister neutron interaction. These abnormal conditions are commonly proposed on Criticality Safety Evaluations. The variety of environments are complemented by the replacement of the SNF for fresh fuel for all the mentioned cases. Through the integrity of this study, a PWR waste management case was considered, including the geometry and composition of the SNF, fuel bundles and canister. 

Considering MCNP as a Monte Carlo calculations reference for OpenMC, a total of $11$ benchmark calculations were conducted, writing equivalent inputs, considering the same: definition of materials, cells and surfaces, execution settings, and cross section library. For comparing the obtained results, a relative difference was used (Equation \ref{eq:diff_per}).

In the NO environment, the k-eff values obtained from OpenMC and MCNP differed by less than $0.8~\%$ for both spent and fresh fuel, with negligible leakage fractions [$0.00001(2*)$], as expected for a large, open geometry. This close agreement suggests minimal inter-canister interaction and demonstrates consistency between the two codes in modeling deep geological repository-like scenarios. While the configurations are based on simplified models, they reflect representative conditions and provide valuable benchmarks for assessing OpenMC’s behavior in realistic SNF storage environments. Additionally, replacing spent fuel with fresh fuel resulted in a $46~\%$ increase in k-eff in both codes, confirming consistent sensitivity to fuel composition.

At the FLD cases, code differences stayed up to $0.8$\% for subcritical systems and $0.55$\% for supercritical systems, replacement of spent fuel for fresh fuel meant a $35$ and $\sim\!40$\% increase in k-eff for cases with and without water leakage into the canister, respectively. Code-to-code k-eff difference resulted $0.7$\% for the periodic lattice calculation. This study considers only one distance between canisters with periodic boundaries, therefore it is suggested further studies with a variable distance in order to look for a potential scenario with wider k-eff differences.
The use of vacuum boundaries in the FLD scenarios resulted in leakage fractions as high as $0.49$, indicating that the calculated k-eff values may be slightly underestimated due to particle loss at the system boundary. To explore the magnitude of this effect, the configuration with the highest leakage was re-simulated using periodic boundary conditions. The resulting k-eff increased by $1.27(2)~\%$ in OpenMC and $1.32(2)~\%$ in MCNP, suggesting that both codes respond similarly to changes in boundary conditions. These findings imply that in scenarios with lower leakage, the impact of boundary effects on k-eff is likely to be smaller. This reinforces the importance of considering boundary treatment when comparing simulation results across codes, particularly in geometries with non-negligible leakage.

In several cases, OpenMC reported an uncertainty of zero for the leakage fraction, which is not physically meaningful given the stochastic nature of Monte Carlo simulations. To address this, a minimum uncertainty of $2$~pcm was adopted. This value corresponds to the smallest detectable fraction $(1/50,000)$ based on the simulation setup, which used $50,000$ particles per batch. This adjustment ensures a conservative and consistent treatment of the leakage fraction, avoiding the misleading impression of absolute certainty and allowing for more meaningful comparisons between cases and between codes.

\begin{figure*}[htb]
    \centering
    \includegraphics[width=0.6\linewidth]{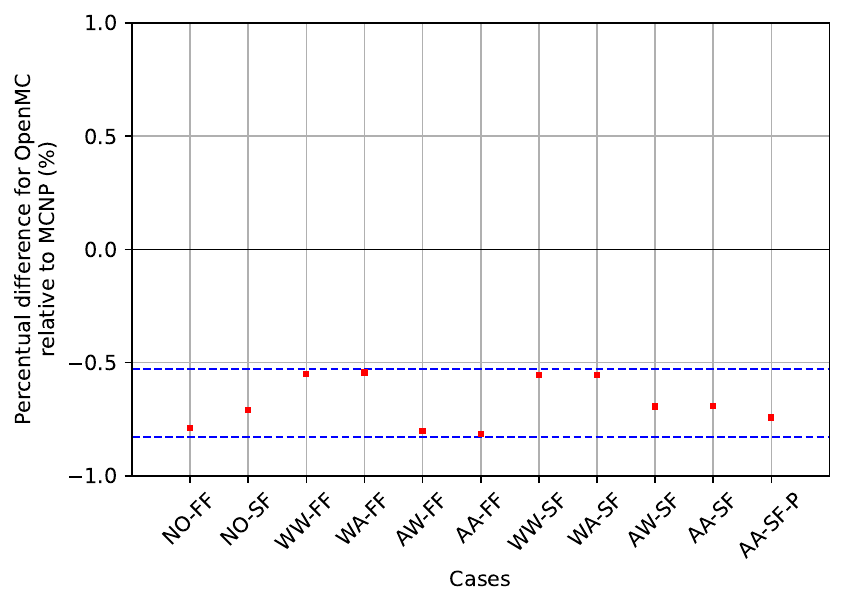}
    \caption{Summary of obtained results. In red, percentual difference relative to MCNP results. In blue, statistical $99.95~\%$ confidence interval. NO: Normal operation, WW: Water in - Water out, WA: Water in - Air out, AW: Argon in - Water out, AA: Argon in - Air out, FF: Fresh fuel, SF: Spent fuel, P: Periodic boundary.}
    \label{fig:summary}
\end{figure*}

A summary of the main results obtained is shown in Figure \ref{fig:summary}. Relative differences present a mean of $-0.682$\% and a standard deviation of $0.109$\%, which supports a tendency of OpenMC to show lower multiplication factors than MCNP.

The safety of a spent fuel storage facility, and of the spent fuel stored within it, is ensured by maintaining subcriticality, besides other functions~\cite{IAEA_storage}. KBS-3 stands as one of the main methods for long term disposal of spent nuclear fuel~\cite{IAEA_status} and criticality safety studies have been made related to this method that support its effectiveness~\cite{Anttila_2005_fit}. Validation of accessible predictive methods should be promoted seeing countries without legal paths to mainstream Monte Carlo codes that have been fully validated through standards like the ICSBEP Handbook. Criticality calculations for systems with spent nuclear fuel contribute to a potential new branch of OpenMC applications, helping to its dependability, while including benchmark methodologies with MCNP supplies code-to-code validation of the obtained results.

\begin{sloppypar}
The differences observed across all benchmark cases remained within $0.9~\%$, indicating a consistent level of agreement between OpenMC and MCNP. OpenMC systematically yielded slightly lower k-eff values than MCNP, with all relative differences falling below zero, suggesting a possible underestimation trend relative to MCNP in these configurations. These results support the potential of OpenMC as a reliable tool for modeling spent nuclear fuel systems, particularly in contexts where open-source accessibility is a priority. To further expand OpenMC’s applicability to regulated scenarios, future comparisons against experimental benchmarks involving spent fuel configurations are recommended. These findings support the ongoing efforts to expand the validation envelope of OpenMC toward dry storage and disposal applications of spent nuclear fuel.
\end{sloppypar}

\section{Data Availability Statement}
The input files used to reproduce the results presented in this study are openly available in Zenodo at \href{https://doi.org/10.5281/zenodo.15749638}{https://doi.org/10.5281/zenodo.15749638}. These files correspond to the \textit{Normal Operation – Spent Fuel} configuration and include both OpenMC and MCNP input decks.

\section{Acknowledgements}
\label{sec:acknowledgments}
J. Romero-Barrientos and F. Molina acknowledge support from ANID FONDECYT Iniciaci\'on Project $11240321$ and ANID FONDECYT Regular Project $1221364$. F. Molina and J. Romero-Barrientos also acknowledge support from ANID - Millennium Science Initiative Program - ICM$2019$\_$044$. 

\section{Declaration of generative AI and AI-assisted technologies in the writing process}
During the preparation of this work, the authors used ChatGPT (developed by OpenAI) to assist with English grammar revision and preparation of ancillary documents such as the README and metadata files for data publication. After using this tool, the authors reviewed and edited the content as needed and take full responsibility for the content of the publication.
\balance
\bibliography{REFS}
\end{document}